\documentclass[twocolumn,showpacs,titlepage,aps,prd,letter,tightenlines,amsmath,byrevtex,nofootinbib,floatfix]{revtex4}

\usepackage{graphicx}

\begin{document}

%%%%%%%%%%%%%%%%%%%%%%%%%%%%%%%%%%%%%%%%%%
\title{High Energy Neutrino Physics with Liquid Scintillation Detectors }
%%%%%%%%%%%%%%%%%%%%%%%%%%%%%%%%%%%%%%%%%%

\author{John G. Learned}
\email{jgl@phys.hawaii.edu}
\affiliation{Department of Physics and Astronomy\\ University of Hawaii, Manoa, 96822 USA}

\date{22 February 2009}

\vglue 1.6cm

\begin{abstract}

Large liquid scintillation detectors have been generally dedicated to low energy neutrino 
measurements, in the MeV energy region (as for example, KamLAND and Borexino). Herein we 
describe the potential employment of large detectors ($>$1 kiloton) for studies of higher 
energy neutrinos interactions, from the cosmic rays and as a long baseline neutrino detector.  
Generally when people have considered large new instruments such as Hanohano and LENA, they 
have abandoned the possibility of doing useful measurements with higher energy neutrino 
interactions since these produce enough light to illuminate every photomultiplier tube (PMT), 
and the scintillation light is isotropic.  Here we take into account Fermat's principle, 
which tells us that indeed the first light to reach the PMTs will be on or near the 
lightcone, the ``Fermat surface'', and that directional track information is available.  
Moreover we have realized that particle type distinction is possible (quasi-elastic muons 
from electrons).  In fact the resolution from a detector of comparable size to 
SuperKamiokande, may be better in both angle, energy and possibly particle type.  This 
realization opens the doors to a number of applications. Moreover, this capability can be 
demonstrated with the use of (future) KamLAND detected events in the new long baseline 
neutrino beam from the Jaeri accelerator in Japan, due to start operations this year.  Some 
of the most attractive possibilities for the future may be in using Hanohano as a moveable 
long-baseline detector in this same beam, and in the employment of LENA in Europe in future 
long baseline neutrino beams from CERN.  Moreover, in the decision as to what type of 
detector to use in a future long baseline experiment from Fermilab to the DUSEL underground 
laboratory at Homestake, consideration should be given to a huge liquid scintillation 
detector.

\end{abstract}


\maketitle

\section{Introduction: the Idea}
\label{introduction}

Liquid scintillation based neutrino detectors have now been employed in sizes up to one kiloton, in 
the KamLAND experiment in Japan\cite{kamland}.  This instrument has been very successful in 
detecting the inverse beta decay interaction caused by electron anti-neutrinos which arrive, after 
neutrino mixing, from reactors around Japan.  The Borexino experiment in Italy has also measured 
neutrinos from the sun, detecting the long sought Boron line.  This physics involves neutrino 
events in the energy range of a few MeV.  The light output of a scintillation detector is 
approximately 250 photoelectrons/MeV (PE/MeV) of deposited ionization energy in KamLAND.  The light 
is considered to be isotropic from near the interaction, and one can only fit a time (to 
nanoseconds), energy (to 5-6\%), and rough location (to 10-20 cm) for such events.  No event 
directionality is obtained. \footnote{Though some directionality may be obtained in a statistical 
way via the direction between positron annihilation and neutron capture in inverse beta decay.  
This was demonstrated by the Chooz Collaboration some years ago.}

The largest underground neutrino detector is the water Cherenkov detector (WC) based 
SuperKamiokande, also in Japan\cite{superk}.  While the energy threshold can be fairly low, perhaps 
4.5 MeV, the statistics are relatively meager at 10 PE/MeV, despite 13,000 PMTs and 40\% coverage 
of the walls of the cylindrical vessel.  The beauty of the Cherenkov technique is however, that one 
has directionality even at low energies.  This has been exploited marvelously by the SuperK group 
to observe the elastic scattering of electrons by incoming solar electron neutrinos, and these 
point sharply back towards the sun.  But this low energy capability still has not been enough to 
detect the low energy events such as the neutron capture on hydrogen which takes place in the 2.2 
MeV delayed event, for identifying electron anti-neutrinos.

So, it has come about that people have accepted that water Cherenkov detectors are needed for 
higher energy neutrino studies and liquid scintillator instruments are deemed superior for lower 
energies.\footnote{ There have been proposals to add water soluble materials (Gadolinium) to 
SuperKamokande in order to make neutrons in inverse beta decay detectable.  At the moment these 
plans are not progressing due to the corrosive nature of the Gadolinium salts. Even so, the 
electron anti-neutrino events below about 5 MeV are not likely to be accessable to SuperK, thus 
missing most of the reactor spectrum and essentially all of the geo-neutrinos.} As many people have 
seen, the SuperKamiokande cosmic ray neutrino interactions make very nice intuitive pictures with 
Cherenkov rings projected on the walls.  Single clean rings result from the most frequent events 
due to $\simeq$1 GeV quasi-elastic muon neutrino (and anti-neutrino) interactions.  Fuzzy, shorter 
rings, more filled-in, result from electron initiated events of the same energy.  Moreover, when 
pi-zero events are present, the pair of decay gammas often produce clearly distinct overlapping 
rings.  In the search for nucleon decay, for example, the back-to-back signature of $p \rightarrow 
e^+ \pi^0$ may be easily recognized.  Of course for higher energy neutrino interactions of a few 
GeV, life becomes more complicated, and one often cannot resolve much detail in the interactions 
(SuperK generally treats such as one single category called "multi-ring events").  This particle 
recognition capability has been very profitable for SuperK, and has resulted in the definitive 
discovery of muon neutrino oscillations in 1998, setting off the neutrino research gold rush of the 
last few years.  New detectors in the megaton class are being considered for Japan, the US and 
Europe.  In the following we will suggest that one may think about an alternative, or supplement, 
to consider very large scintillation detectors as well.

 %%%%%%%%%%%%%%%%%% FIGURE 1 %%%%%%%%%%%%%%%%
\begin{figure}[htbp]
%\vglue -3.0cm
\begin{center}
%\hglue -0.5cm
\includegraphics[width=0.5\textwidth]{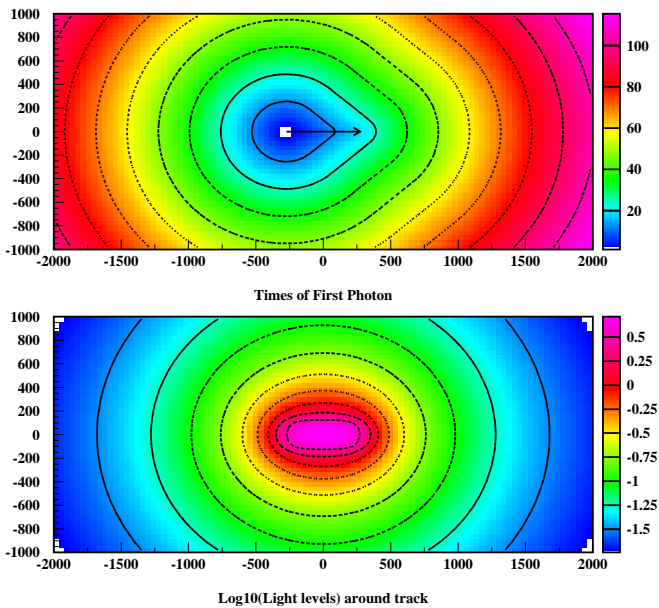}
\end{center}
%\vglue -0.4cm

\caption{These panels illustrate the time of first photon arrival (upper) and light level 
(lower) for a 547 cm long "muon" track in liquid scintillator in a 20m x 40m detector. One sees 
the asymmetric time contours due to photons on the Cherenkov cone. Times point to the start 
of the track, while amplitudes reveal the track center. }

\label{fig:tqcont} 
\end{figure}
%%%%%%%%%%%%%%%%%% FIGURE 1 %%%%%%%%%%%%%%%%%

\subsection{Scintillation Detectors Really are also Cherenkov Detectors for Higher Energies}

 Imagine a quasi elastic muon neutrino interaction in something like Hanohano or even 
KamLAND (or SNO+ or LENA), a single track of a few meters length.  All researchers I know 
(including me) have been saying that all we would get is a nice calorimetric 
measurement\footnote{Through-going muons are fitted by virtue of the evident entry point 
(``wound'') in the response of the PMTs adjacent to the entry and exit points. Sanshiro 
Enomoto and Dan Dwyer of KamLAND have employed some timing of this type to fit 
through-going muons.}. One could measure total energy well, but nothing of direction or 
event topology.  The realization herein (obvious {\it a posteriori}) that if one employs 
the earliest arriving hits on PMTs these will be on the light cone, and hence (by Fermat's 
Principle) will indeed have the timing as for a Cherenkov cone. Either side of the 
Cherenkov cone the light timing will be as from a point source, so it is this combination 
of partial spherical and cylindrical radiation that forms the ``Fermat Surface''. The 
greater the light, the better will be the approximation that the early hits are close to 
the Cherenkov time. For SuperK filled with liquid scintillator (and other possible 
detectors we will discuss) this would be hundreds of PE per tube for a 1 GeV event, and the 
earliest arrival will typically be within about one nanosecond of the lightcone. As is well 
known, this cone has wonderful directivity.\footnote{One concern for this first-in PE 
analysis is making sure that the PMTs have a low rate of prepulsing.  Fortunately recent 
generations of PMTs do have very low rates of prepulsing.  However, even if there is some 
prepulsing it can be dealt with through neighbor algorithms.}

\subsection{There is plenty of light for first-in hit fitting}

The trick here is that one can, with the relatively large event energies (say, GeV compared 
to MeV threshold sensitivity), get a hit nearly at the earliest possible time in almost 
every phototube.  For example, if we take the sensitivity of KamLAND one would expect a 
total of 250PE/MeV x 1000MeV / 2000PMTs = 250 PE/PMT for a 1 GeV muon.  (KamLAND has about a 
20\% photocathode coverage, and about a 20\% quantum efficiency.)  A 1 GeV muon will travel 
for a distance of roughly 500 cm, so on average the tubes will get about 1 PE from each 2 cm 
equivalent track segment (assuming isotropic radiation of photons from the track, as is the 
case).  The geometric arrival time is spread over a time of order less than 60 ns, but 
somewhat more spread out due to scattering, absorption and re-emission, etc.  Note that 
because of Fermat's Principle, no photon can beat the light cone, only fall behind it.

\subsection{Real Cherenkov radiation will make some contribution}

Depending upon the geometry, energy and scintillator, Cherenkov radiation may make a useful 
prompt light boost within the (small) solid angle of the tracks on the light cone.  For an 
order of magnitude estimate, the equivalent Cherenkov radiation in KamLAND should be about 
40 PE/cm * 0.2 coverage * 500 cm = 4,000 PE which will land in the tubes illuminated by the 
Cherenkov cone, which I would take to be for this case to be about 1/4 on average. This is 
to be compared to roughly 60,000 PE from the scintillator light in the cone.  Hence, very 
roughly I would expect that the in-cone Cherenkov signal will nearly double the first-in 
PEs from the scintillator.  This is terrific:  we can have a very good directional 
signal... we should be able to fit the direction of a quasi-elastic single muon quite 
nicely, as we demonstrate below.  We did not include Cherenkov radiation in the simulations 
described herein in order to be conservative, but also there are complications in modeling 
the direct (and very blue) Cherenkov light which will be somewhat absorbed (and some 
re-emitted shifted towards the red).  So, how much this improves the real world situation 
remains to be studied.

\subsection{Other methods of event topology resolution}

With this amount of light available it may well be profitable to install some cameras along 
with the photomultipliers.  One may consider a small scale detector in which the light 
collected by commercial cameras would be substantial (eg. meter scale detectors observed 
with $\simeq$10 cm lenses, possibly useful for reactor montoring).  However, for the very 
large instruments we focus upon herein, apertures on the scale of a substantial fraction of 
1 m$^2$ would be needed.  I will not pursue this line of thought here, but one may think 
about something like a version of the RICH detectors discussed by Tom Ypsilantis, with a 
spherical mirror and where the (spherical) focal surface is occupied by imaging or 
pixelated PMT.  Pieces of Cherenkov rings are imaged as ``moustache'' shapes, while point 
sources will form point images, and a point moving across the field of view will image to a 
line.  Depth of field is a problem, but solvable with multiple views (I 
believe).\footnote{I proposed using CCD cameras to image events in KamLAND at the 
collaboration in April 2005.  The idea was partly to turn the imaging device invented by 
Makoto Sasaki for UHE cosmic ray work around and look inwards. Some simulation work was 
done by Byron Dieterle, and experimental tests have been conducted by groups at Tohoku 
University and at Kansas State university.  Other work is progressing along these lines for 
directional low energy neutrino studies in various agencies.  But until now, as far 
as I know, all this has been aimed at the few MeV energy regime.}

\section{Geometry}

Imagine a single muon moving horizontally (take it to be +x direction, rightwards) through 
the middle of a recumbent cylindrical detector (40 m long by 20 m in diamter), and which 
track extends some 5 m meters, starting at x = -2.5m and travels to x = +2.5m.  As the muon 
progresses it will produce lots of light, as said, but let us only consider the leading 
light.  That light will be on the Cherenkov cone heading out in y and z, and heading off 
diagonally to the right (+x).  Projected back to the track these first hit times will fit 
up, just as is done for Cherenkov cones in SuperK.  But one will get something else too.  
Those PMTs to the left of the illumination zone of the Cherenkov cone will get hits from 
near the start of the track, and will fit to a nice spherical point-fit.  Similarly those 
PMTs to the right, after the end of the track and Cherenkov cone, will also later receive 
hits on a spherical front centering on the end of the track.  Hence, in principle, in this 
case (and every case for a single straight track) one can fit the all leading times with 
two point fits plus a Cherenkov cone, the Fermat Surface, as illustrated in Figure 
\ref{fig:tqcont}. 

%%%%%%%%%%%%%%%%%% FIGURE 2 %%%%%%%%%%%%%%%%
\begin{figure}[htbp]
%\vglue -3.0cm
\begin{center}
%\hglue -0.5cm
\includegraphics[width=0.5\textwidth]{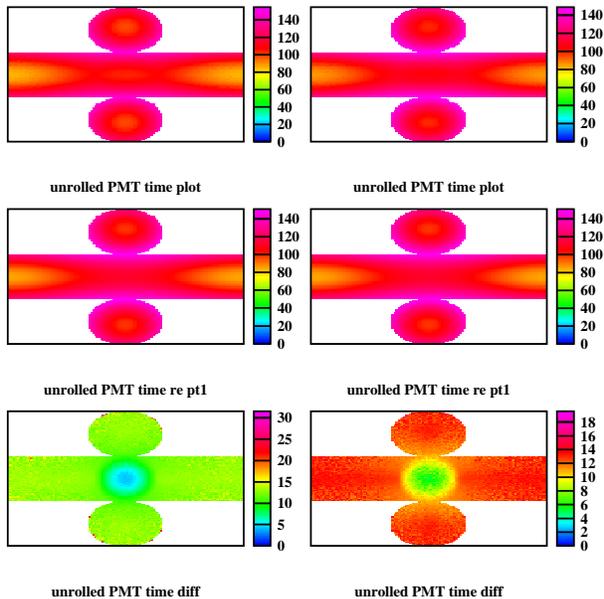}
\end{center}
%\vglue -0.4cm

\caption{These panels show first arrival times, for a muon on the left and and electron 
shower on the right.  The top panels are the simulated data for a 1 GeV event (for the muon a 
574 cm m track).  The second row show times and amplitudes if the signal was from the 
interaction vertex.  The lower two panels show the residuals between simulated data these and 
point fits, illustrating the difference between a muon event and an electron induced event}

\label{fig:tptfits} 
\end{figure}
%%%%%%%%%%%%%%%%%% FIGURE 2 %%%%%%%%%%%%%%%%%

\begin{figure}[htbp]
%\vglue -3.0cm
\begin{center}
%\hglue -0.5cm
\includegraphics[width=0.5\textwidth]{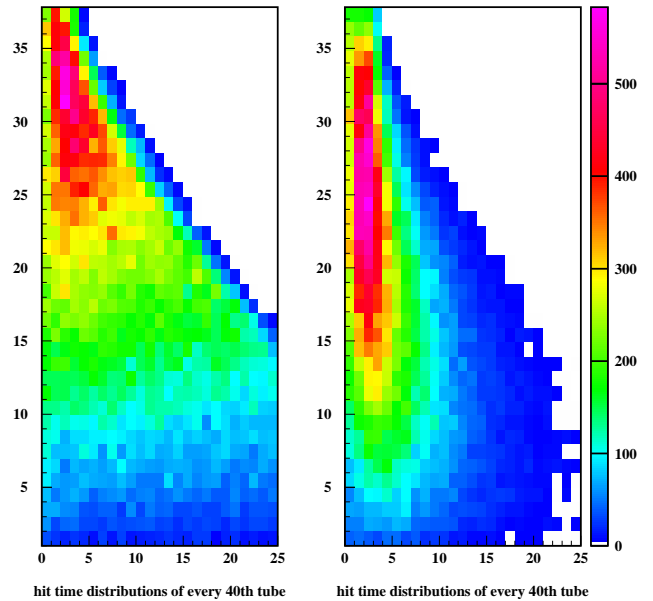}
\end{center}
%\vglue -0.4cm

\caption{A selection of individual time distributions for phototubes on the top of 
the 40 m x 40 m cylindrical detector studied, for a 1 GeV horizontal muon track and 
similar electron shower.  The times are relative to the first photon for that 
phototube.  One sees very significant differences in these distributions, and hence 
that there is much information to be potentially extracted from the individual 
photomultiplier record of pulse shape.  Of course these are idealized plots with no 
optical scattering and such.}

\label{fig:tdists} 
\end{figure}

Do we have enough information?  Of course, there are a few thousand illuminated PMTs each with a time 
and a total charge, and, most probably in a new detector, full waveform records.  The fit is greatly 
overconstrained (7 or 8 unknowns and several thousand data points!).  But this is not to prove that the 
information is not so smeared as to be impossible to extract.  For example, the Fermat Surface (the 
equi-time countours, not to be confused with amplitude) represents the times of the nearest tracks, as 
in the case of multiple tracks from the same vertex. Only those track segments nearest the detector 
surface will register the time evolution for said track.  Hence, for example, a quiver of tracks 
emanating from the same vertex point will only reveal the bounding track elements in the Fermat 
Surface.  A track in the bunch center will be obscured.  Hence we can conclude that from leading 
timing, no complex multi-track topology will be revealed.  Of course, using the total light and hence 
energy deposited we may know roughly what was going on and a volume within which it was constrained, 
but no bubble chamber-like images can be extracted.  (Which is also true of Cherenkov recordings as in 
SuperK of multi-particle final states).  Yet, there is further information, buried in the multiple hits 
on each PMT.  I hestitate to claim how well this can be used, since here will also lie all the 
complications of optical scattering, re-emission and so on.  None-the less, in Figure \ref{fig:tdists}, 
one sees that there are gross differences between the photon arrival times for a muon track and for the 
same energy (1 GeV) electron shower.

%%%%%%%%%%%%%%%%%% FIGURE 3 %%%%%%%%%%%%%%%%
%%%%%%%%%%%%%%%%%% FIGURE 3 %%%%%%%%%%%%%%%%%

In order to explore this I have written a small simulation program.  The pretend detector 
is a right cylinder 40 m in diameter and 40 m high, 50,265 $m^3$ and has a fill of next 
generation liquid scintillator (doped linear alki-benzene, LAB, density 0.86) producing on 
average 10,000 detectable photons/MeV.  I take it to have 7,377 phototubes each with a 
collection area of 0.4 m$^2$, mounted on a 102 cm lattice.  I ignore optical scattering and 
attenuation (not important for present purposes).  In the program the muon is stepped long 
at the speed of light in vacuum while radiating 20,000 photons per cm, in random 
directions.  Each photon is tracked towards the detector wall at the speed of light in 
vacuum divided by the index of refraction of the medium (n = 1.482).  A resolution of 2 ns 
is inserted. The time of the first hit in, the total number of hits and the median time are 
saved for each PMT.

%%%%%%%%%%%%%%%%%% FIGURE 4 %%%%%%%%%%%%%%%%
\begin{figure}[htbp]
%\vglue -3.0cm
\begin{center}
%\hglue -0.5cm
\includegraphics[width=0.5\textwidth]{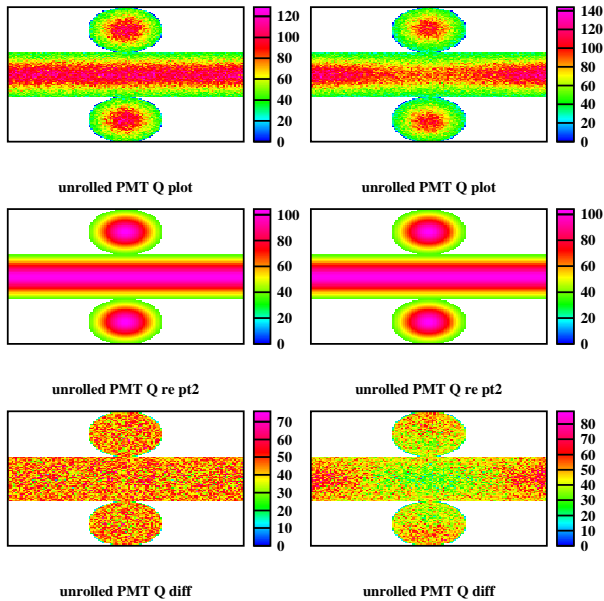}
\end{center}
%\vglue -0.4cm

\caption{These panels show charge collected for a muon event on the left and an electron 
event on the right.  The top panels are the simulated data for 1 GeV events.  The second row 
show amplitudes if the signal was from a point at the middle of the track.  The lower two 
panels show the residuals between simulated data and point radiation, illustrating that point 
fits to the charge measured, reveal the middle of track very well.}

\label{fig:qptfits} 
\end{figure}
%%%%%%%%%%%%%%%%%% FIGURE 4 %%%%%%%%%%%%%%%%%

Exploring this ``data'', I find that, as one may guess, a fit to the total charge collected 
for each PMT gives a nice determination of the track center, as illustrated in Figure 
\ref{fig:qptfits}.  These panels are modelled on the SuperKamiokande unrolled view, with 
top folded up, bottom down, and the barrel of the cylinder unrolled. The simulated muon 
goes through the detector center and is pointed towards the back. The top panel shows the 
charge collected by each PMT, the second panel the charge expected from a point source of 
radiation of the same total amount of light, but radiating from the center of the track.  
The third panel shows the difference between these two data sets, and as one sees the point 
fit gives a very nice reproduction, indicating a simple means to start track finding.  It 
turns out that calculating the ``center of mass'' of the detected charge (or photon levels) 
gives a good measure of the track center, without any fitting.

Figure \ref{fig:tptfits} illustrates the data in terms of first-in photon time. The next 
panel down has the times as calculated from the begining of the muon track.  The lower left 
panel shows the difference between the top two, and such small residuals show that a point 
time fit reveals the track origin fairly well. On the right side the same things are 
presented for a 1 GeV electron event.

\begin{figure}[htbp]
%\vglue -3.0cm
\begin{center}
%\hglue -0.5cm
\includegraphics[width=0.5\textwidth]{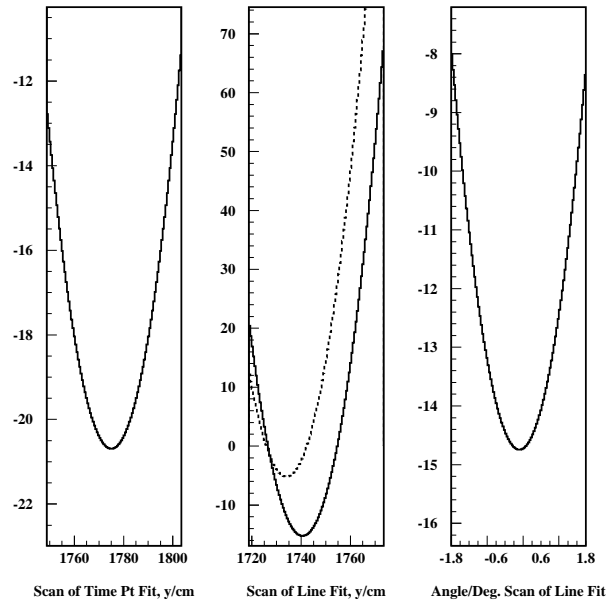}
\end{center}
%\vglue -0.4cm

\caption{Resolutions for point fits to time, at start of track; line to whole 
track; and varying angle of track. The line-fit scan shows the fit to an electron 
shower, on the same scale, which is about 11 $\sigma$ equivalent worse.}

\label{fig:res} 
\end{figure}

I have tried a time based point fit, moving the test point, and this yields a point 
close to the beginning of the track (with a resolution of $<$10 cm, though 
systematically shifted by 80 cm, as illustrated in Figure \ref{fig:res}.  These 
fits are very useful as a method to find the first guess at a track.  With 
development one can probably do better, but this does point towards a quick start 
to track finding.

%%%%%%%%%%%%%%%%%% FIGURE 5 %%%%%%%%%%%%%%%%
%%%%%%%%%%%%%%%%%% FIGURE 5 %%%%%%%%%%%%%%%%%

The third fitting made was to the Fermat timing of a line (pseudo muon), defined by 
a starting and an ending point. A little geometric derivation yields the formula 
that the time from the start of the track to a point given by the vector ${\bf 
R_p}$, where the unit vector ${\bf R_t}$ points long the track and the angle 
between track and PMT direction (given by ${\bf cos(\alpha) = R_p R_t/|R_p|}$) is:  
$\bf{ ct = {|R_p|} [sin(\alpha) tan(\theta_C) +cos(\alpha)]}$.  This makes for fast 
fitting, and some results are shown in Figure \ref{fig:res}, two right hand panels, 
which show a $\chi^2$ scan as a test track is moved along the `real' "track and as 
the test track is varied in angle with respect to the real "track.  The results 
encourage: the resolution in longitudinal fitting is about 1 cm, and in angle in 
about 0.2 degrees.  Now perhaps these numbers are a bit optimistic, but it is clear 
that they are competitive with and perhaps even better than those for SuperK. (The 
SuperK-I single ring resolution for e's was 3 deg, $\mu$'s 1.8 deg\cite{superk}).

\section{Excellent Energy Resolution Too}

If we collect 250,000 PE for a 1 GeV event, then the energy resolution could be as good as 
0.2\% for each event based upon Poisson statistics.  Of course calibrations, non-linearities 
and such systematics will spoil this a bit, but it still will likely compare favorably with 
the SuperK momentum resolution for e's of 3.2\% and 2.5\% for $\mu$'s at 1 
Gev/c\cite{superk}.  But is this possible advantage useful?  It does mean that for events in 
which we know the source direction, as in T2K from the accelerator (or in LENA from CERN, or 
Hanohano as a long baseline experiment in the water near Korea), we may measure the energy 
and the angle relative to the direction towards the acccelerator well, thus knowing the 
original neutrino energy, hence the L/E, whence facilitating precision measurements of muon 
neutrino oscillations.  For comic ray neutrino studies the muon direction gives a conical 
locus of directions for the incoming neutrino, which is not as useful. 

Since the fit of the initial point radiation (on the left in the situation described above 
and pictured in Figure \ref{fig:tqcont}) will yield the light near the start of the track we 
have the exciting prospect to deduce the energy of the nuclear recoil from the relative 
number of the early hits.  Of course the recoil nucleon will suffer from Birks saturation of 
the scintillator in its short energy deposition region, but clearly this is another area for 
study.

\section{Flavor Discrimination}

In order to accomplish further interesting neutrino physics study it would be wonderful to be 
able to discriminate electron generated showers from muon tracks.  We know the terrific 
success of IMB/Kam/SuperK in doing just this, simply based upon the fuzziness and extent of 
the Cherenkov cone.  The EM shower builds quickly and flares towards the end and with 
increasing energy deposition (on average the longitudinal energy deposition profile may be 
approximated by a power law times a Gaussian), while the muon track energy deposition is more 
constant.  Since the EM shower has more particles this effect somewhat fills the cone.  But 
principally the electron shower is shorter as we move to higher energies: only a meter or so 
for electrons, compared to about  five meters at one GeV for a muon. The electron shower 
may also reveal itself through the textural fluctuations in the Fermat Surface towards it's 
tail.

The SuperK single ring electron detection efficiency is 90.2\% and muon 95.8\%, but 
most importantly only 0.4\% of the muons are identified as electrons\cite{superk}. 
Fitting the simulated muon event to a line radiation yields about an eleven standard 
deviation (11 $\sigma$) better fit than the same test function to a same energy 
simulated electron shower distribution. More careful simulation is assuredly needed 
here, and though the $e/\mu$ separation will deteriorate with lower energies and with 
inclusion of realistic optics, the author's sense is that the liquid scintillation 
detector will be quite competitive.

There may be some advantage as well, since for purposes of electron neutrino appearance, 
the most annoying background process is the single neutral pion production in the case that 
one of the two decay gammas goes nearly straight ahead and the second is unobserved.  The 
scintillation detector might pick up either indications of the backwards photon and/or the 
recoil nucleon.  However, this suggestion is assuredly speculative, and the author 
understands that work is needed.

I have made some preliminary investigations of flavor discrimination. One promising avenue 
is calculating the moment of inertia of the charge, weighted by time.  The eigenvectors 
give a good first try at direction, and I find that the results for the same energy muon 
and electron are quite different (opposite directions), and this is with no fitting. I 
conclude that flavor separation will be good in the few hundred MeV to several GeV range.  
How good remains to be determined.

\section{Possible Applications}

It would appear that this type of detector may have some significant applications, aside 
from the expected ones extrapolating from experience with KamLAND and SuperK employing 
studies of geoneutrinos, reactor neutrinos, relic supernova neutrino searches, remote 
reactor monitoring, etc.

\subsection{T2KL}

There is much effort going into the T2K, or Tokai to Kamoka, experiment, with a predicted 
rate of 3,900 events per year in 22.5kT SK for the first T2K configuration\cite{t2k}.  This 
may go up by a factor of 5 with an increased power accelerator beam later.  If we just scale 
by factor of 1/22.5 from SK to KamLAND there should be roughly 20 events per year in KL, or 
100 events in five years (assumed for a T2K run).  (This was for a 100 day year and 2 deg 
off axis\cite{t2k}).  So, does this give hope to do some good physics with KamLAND in the 
T2K beam?  Having the rate down by 22.5x certainly hurts, leaving KamLAND uncompetitive for 
any $\theta_{13}$ work.  Is there some advantage one may identify here?  I suspect that 
there may be a good $\theta_{23}$ measurement (recall that the accuracy here has to 
do with fitting the period of oscillation, and errors are not simply Poissonian). 

Of course it is always good to look for unusual physics... though I do not readily think of 
what that might be.  Perhaps there is some physics which can be explored employing the 
incoming and uncontained events (where KamLAND is only down by a factor of six or so from 
SuperKamioande)?

In any event, KamLAND can assuredly provide a (free!) test bed for a much larger liquid 
scintillator experiment. The neutrinos will be there and presumably the signals collected.  
Timing will select true signature from background so whatever is seen should be clean of 
cosmic ray background.

\subsection{Hanohano and LENA Long Baseline}

One obvious application of large scintillation detectors comes immediately to mind: perhaps 
one may employ Hanohano in a long baseline run from Tokai, located in the water between 
Japan and Korea and forming a second location beyond Kamioka (and perhaps a third will be 
built in Korea).  At 10,000 tons we would get about half the rate, and at those the distance 
it would fall to 1/8 the rate of SK.  This is still substantial, though larger would be 
better if HyperKamiokande is going to be built.  Studies for Hanohano have indicated that 
mobile ocean based liquid scintillation detectors up to about 100 kilotons are practical 
from an engineering standpoint\cite{hanohano}. Note that for this application one does not 
need to be very deep (as is the case for geoneutrino studies, for example which require 3 km 
or more of water overburden).  The relatively large accelerator neutrino energies and the 
small duty cycle of the accelerator permits strong rejection of the cosmic ray background.  
Perhaps an alternative location would be in Lake Superior as a long baseline station from 
Fermilab, useful perhaps with Nova.

Similarly, the proposed LENA detector in Europe, at 50 kilotons, may now be considered as an 
end station for future CERN long baseline neutrino beams (in competition with a proposed few 
hundred kiloton water Cherenkov detector and a hundred kiloton liquid argon instrument).

\subsection{WIMP Indirect Detection}

 For energies in the same 1 GeV range we are talking about, a large liquid scintillation 
instrument will certainly provide very nice detection of neutrinos.  For indirect WIMP 
detection one should expect flavor democracy (plenty of time to equilibrate), and these 
events should point to the sun.  If we really can get 0.2 degree angular resolution that 
would be terrific, but 1 degree would do.  However the weak interaction scattering angle 
spoils nice pointing: having good angle determination for the muon does not help if the weak 
scattering angle is 20 degrees.  Yet, having good energy and angle measurement means that the 
neutrino energy of any events from the solar direction will be well determined.  Hence if we 
are so lucky as to have substantial WIMP annihilation to a two neutrino mode then this method 
may prove decisive in resolving the signature from the cosmic ray neutrino background.  
Again, this needs more work but it seems attractive, at least in the instance that the WIMP 
mass is not in the SUSY favored 100 GeV range, but rather in the few GeV range.

\subsection{Nucleon Decay}

 There have been suggestions for some time to employ employ a large liquid scintillation 
detector for observing nucleon decay (for example with LENA\cite{lena}), particularly in the 
instance of SUSY favored modes involving a kaon which has kinetic energy too low to make 
Cherenkov light.  With the present realization of the ability to study the directionality, 
the prospects are certainly improved.  In order to compete with SuperK, probably a 100 
kiloton scintillation detector would be needed.

\subsection{Astrophysics}

 Once again, there have been many searches for astrophysical phenomena discussed for smaller 
instruments and instruments with low thresholds. Aside from the WIMP search mentioned above, 
and extending searches as explored by KamLAND (relic supernova neutrinos, anti-neutrinos 
from the sun, bursts of low energy neutrinos from such sources as AGNs, micro-quasars, etc., 
a large liquid scintillation detector can enlarge the search for astrophysical neutrinos in 
the GeV range.

\section{Summary and Outlook}

 The notion to employ a large liquid scintillation instrument as a detector which can 
accomplish similar physics to that which can be done with a large water Cherenkov detector 
has been sketched out.  In some ways, for the GeV scale events, such a detector may equal 
or outperform the water Cherenkov instruments. Certainly liquid scintillation based 
instruments have the advantage of simultaneously being able to explore the lower energy 
range of electron anti-neutrinos due to reactors and geological origins, which are 
inaccessible to a water based Cherenkov detector. At the very least, I think we can 
conclude that this line of large neutrino detector consideration deserves further efforts.

\section{acknowledgements}

 The idea in this note came out during conversations with Jason Kumar, and subsequent 
discussions with Shige Matsuno, Mikhael Batygov, Steve Dye and Sandip Pakvasa.  Thanks to 
all.  Also thanks to the AMANDA team whose use of first-in hits in fitting events in ice 
where the photon propagation is diffusive, probably put me in mind of this idea.  Also thanks 
to Bob Svoboda who some years ago suggested converting SuperK to a liquid scintillation 
detector in his SCIPIO plan.  And thanks to Sanshiro Enomoto and Dan Dwyer for their thesis 
work on through-going muon fitting in KamLAND.

\end{document}